\newif\ifPAGELIMIT
\PAGELIMITfalse
\documentclass[conference,letterpaper]{IEEEtran}

\usepackage[numbers,sort&compress]{natbib}

\makeatletter

\usepackage{etex}

\usepackage{zref-savepos}

\newcounter{mnote}

\def\xmarginnote{%
  \xymarginnote{\hskip -\marginparsep \hskip -\marginparwidth}}

\def\ymarginnote{%
  \xymarginnote{\hskip\columnwidth \hskip\marginparsep}}

\long\def\xymarginnote#1#2{%
\vadjust{#1%
\smash{\hbox{{%
        \hsize\marginparwidth
        \@parboxrestore
        \@marginparreset
\footnotesize #2}}}}}

\def\mnoteson{%
\gdef\mnote##1{\refstepcounter{mnote}\label{##1}%
  \zsavepos{##1}%
  \ifnum20432158>\number\zposx{##1}%
  \xmarginnote{{\color{blue}\bf $\langle$\arabic{mnote}$\rangle$}}%
  \else
  \ymarginnote{{\color{blue}\bf $\langle$\arabic{mnote}$\rangle$}}%
  \fi%
}
  }
\gdef\mnotesoff{\gdef\mnote##1{}}
\mnoteson
\mnotesoff

\usepackage{framed}
\usepackage{subcaption}
\usepackage{comment}
\usepackage[svgnames,dvipsnames]{xcolor}
\usepackage[all]{xy}
\usepackage{tikz}
\usetikzlibrary{positioning,matrix,through,calc,arrows,fit,shapes,decorations.pathreplacing,decorations.markings,}

\tikzstyle{block} = [draw,fill=blue!20,minimum size=2em]

\usepackage{qsymbols,amssymb,mathrsfs}
\usepackage{amsmath}
\usepackage[standard,thmmarks]{ntheorem}
\theoremstyle{plain}
\theoremsymbol{\ensuremath{_\vartriangleleft}}
\theorembodyfont{\itshape}
\theoremheaderfont{\normalfont\bfseries}
\theoremseparator{}

\theoremstyle{nonumberplain}
\theoremheaderfont{\scshape}
\theorembodyfont{\normalfont}
\theoremsymbol{\ensuremath{_\blacktriangleleft}}

\theoremnumbering{arabic}
\theoremstyle{plain}
\usepackage{latexsym}
\theoremsymbol{\ensuremath{_\Box}}
\theorembodyfont{\itshape}
\theoremheaderfont{\normalfont\bfseries}
\theoremseparator{}

\theorembodyfont{\upshape}
\theoremprework{\bigskip\hrule}
\theorempostwork{\hrule\bigskip}

\usepackage[overload]{empheq} 

\let\iftwocolumn\if@twocolumn
\g@addto@macro\@twocolumntrue{\let\iftwocolumn\if@twocolumn}
\g@addto@macro\@twocolumnfalse{\let\iftwocolumn\if@twocolumn}

\mathtoolsset{showonlyrefs=false,showmanualtags}
\let\underbrace\LaTeXunderbrace 

\renewcommand{\eqref}[1]{\textup{(\refeq{#1})}} 
\newtagform{brackets}[]{(}{)}   
\usetagform{brackets}

\usepackage[Smaller]{cancel}

\PassOptionsToPackage{breaklinks,hyperindex=true,backref=false,bookmarksnumbered,bookmarksopen,linktocpage,colorlinks,linkcolor=BrickRed,citecolor=OliveGreen,urlcolor=Blue,pdfstartview=FitH}{hyperref}
\usepackage{hyperref}

\usepackage{graphicx,psfrag}
\graphicspath{{figure/}{image/}} 

\usepackage{diagbox} 

\usepackage{algorithm2e}
\usepackage{listings} 
\lstdefinelanguage{Maple}{
  morekeywords={proc,module,end, for,from,to,by,while,in,do,od
    ,if,elif,else,then,fi ,use,try,catch,finally}, sensitive,
  morecomment=[l]\#,
  morestring=[b]",morestring=[b]`}[keywords,comments,strings]
\DeclareMathAlphabet{\mathpzc}{OT1}{pzc}{m}{it}
\usepackage{upgreek} 
\usepackage{dsfont}  

\def\multi@nostar#1#2{%
  \expandafter\def\csname multi#1\endcsname##1{%
    \if ##1.\let\next=\relax \else
    \def\next{\csname multi#1\endcsname}     
    \expandafter\newcommand\csname #1##1\endcsname{#2}
    \fi\next}}

\def\multi@star#1#2{%
  \expandafter\def\csname #1\endcsname##1{#2}
  \multi@nostar{#1}{#2}
}

\newcommand{\multi}{%
  \@ifstar \multi@star \multi@nostar}

\multi*{rm}{\mathrm{#1}}
\multi*{mc}{\mathcal{#1}}
\multi*{op}{\mathop {\operator@font #1}}
\multi*{ds}{\mathds{#1}}
\multi*{set}{\mathcal{#1}}
\multi*{rsfs}{\mathscr{#1}}
\multi*{pz}{\mathpzc{#1}}
\multi*{M}{\boldsymbol{#1}}
\multi*{R}{\mathsf{#1}}
\multi*{RM}{\M{\R{#1}}}
\multi*{bb}{\mathbb{#1}}
\multi*{td}{\tilde{#1}}
\multi*{tR}{\tilde{\mathsf{#1}}}
\multi*{trM}{\tilde{\M{\R{#1}}}}
\multi*{tset}{\tilde{\mathcal{#1}}}
\multi*{tM}{\tilde{\M{#1}}}
\multi*{baM}{\bar{\M{#1}}}
\multi*{ol}{\overline{#1}}

\multirm  ABCDEFGHIJKLMNOPQRSTUVWXYZabcdefghijklmnopqrstuvwxyz.
\multiol  ABCDEFGHIJKLMNOPQRSTUVWXYZabcdefghijklmnopqrstuvwxyz.
\multitR   ABCDEFGHIJKLMNOPQRSTUVWXYZabcdefghijklmnopqrstuvwxyz.
\multitd   ABCDEFGHIJKLMNOPQRSTUVWXYZabcdefghijklmnopqrstuvwxyz.
\multitset ABCDEFGHIJKLMNOPQRSTUVWXYZabcdefghijklmnopqrstuvwxyz.
\multitM   ABCDEFGHIJKLMNOPQRSTUVWXYZabcdefghijklmnopqrstuvwxyz.
\multibaM   ABCDEFGHIJKLMNOPQRSTUVWXYZabcdefghijklmnopqrstuvwxyz.
\multitrM   ABCDEFGHIJKLMNOPQRSTUVWXYZabcdefghijklmnopqrstuvwxyz.
\multimc   ABCDEFGHIJKLMNOPQRSTUVWXYZabcdefghijklmnopqrstuvwxyz.
\multiop   ABCDEFGHIJKLMNOPQRSTUVWXYZabcdefghijklmnopqrstuvwxyz.
\multids   ABCDEFGHIJKLMNOPQRSTUVWXYZabcdefghijklmnopqrstuvwxyz.
\multiset  ABCDEFGHIJKLMNOPQRSTUVWXYZabcdefghijklmnopqrstuvwxyz.
\multirsfs ABCDEFGHIJKLMNOPQRSTUVWXYZabcdefghijklmnopqrstuvwxyz.
\multipz   ABCDEFGHIJKLMNOPQRSTUVWXYZabcdefghijklmnopqrstuvwxyz.
\multiM    ABCDEFGHIJKLMNOPQRSTUVWXYZabcdefghijklmnopqrstuvwxyz.
\multiR    ABCDEFGHIJKL NO QR TUVWXYZabcd fghijklmnopqrstuvwxyz.
\multibb   ABCDEFGHIJKLMNOPQRSTUVWXYZabcdefghijklmnopqrstuvwxyz.
\multiRM   ABCDEFGHIJKLMNOPQRSTUVWXYZabcdefghijklmnopqrstuvwxyz.

\newcommand{\dotleq}{\buildrel \textstyle  .\over {\smash{\lower
      .2ex\hbox{\ensuremath\leqslant}}\vphantom{=}}}
\newcommand{\dotgeq}{\buildrel \textstyle  .\over {\smash{\lower
      .2ex\hbox{\ensuremath\geqslant}}\vphantom{=}}}

\newcommand{\bM}{\begin{bmatrix}}
\newcommand{\eM}{\end{bmatrix}}
\newcommand{\bSM}{\left[\begin{smallmatrix}}
\newcommand{\eSM}{\end{smallmatrix}\right]}
\renewcommand*\env@matrix[1][*\c@MaxMatrixCols c]{%
  \hskip -\arraycolsep
  \let\@ifnextchar\new@ifnextchar
  \array{#1}}

\newqsymbol{`N}{\mathbb{N}}
\newqsymbol{`R}{\mathbb{R}}
\newqsymbol{`Z}{\mathbb{Z}}

\newqsymbol{`|}{\mid}
\newqsymbol{`8}{\infty}
\newqsymbol{`1}{\left}
\newqsymbol{`2}{\right}
\newqsymbol{`6}{\partial}
\newqsymbol{`0}{\emptyset}
\newqsymbol{`-}{\leftrightarrow}

\DeclareFontFamily{OMX}{MnSymbolE}{}
\DeclareSymbolFont{MnLargeSymbols}{OMX}{MnSymbolE}{m}{n}
\SetSymbolFont{MnLargeSymbols}{bold}{OMX}{MnSymbolE}{b}{n}
\DeclareFontShape{OMX}{MnSymbolE}{m}{n}{
    <-6>  MnSymbolE5
   <6-7>  MnSymbolE6
   <7-8>  MnSymbolE7
   <8-9>  MnSymbolE8
   <9-10> MnSymbolE9
  <10-12> MnSymbolE10
  <12->   MnSymbolE12
}{}
\DeclareFontShape{OMX}{MnSymbolE}{b}{n}{
    <-6>  MnSymbolE-Bold5
   <6-7>  MnSymbolE-Bold6
   <7-8>  MnSymbolE-Bold7
   <8-9>  MnSymbolE-Bold8
   <9-10> MnSymbolE-Bold9
  <10-12> MnSymbolE-Bold10
  <12->   MnSymbolE-Bold12
}{}

\let\llangle\@undefined
\let\rrangle\@undefined
\DeclareMathDelimiter{\llangle}{\mathopen} {MnLargeSymbols}{'164}{MnLargeSymbols}{'164}
\DeclareMathDelimiter{\rrangle}{\mathclose} {MnLargeSymbols}{'171}{MnLargeSymbols}{'171}

\DeclarePairedDelimiter\Span{\llangle}{\rrangle}
\DeclarePairedDelimiter\abs{\lvert}{\rvert} 

\DeclarePairedDelimiter\ceil{\lceil}{\rceil}
\DeclarePairedDelimiter\floor{\lfloor}{\rfloor}
\DeclarePairedDelimiter\Set{\{}{\}}
\newcommand{\imod}[1]{\allowbreak\mkern10mu({\operator@font mod}\,\,#1)}

\newcommand{\threecols}[3]{
\hbox to \textwidth{%
      \normalfont\rlap{\parbox[b]{\textwidth}{\raggedright#1\strut}}%
        \hss\parbox[b]{\textwidth}{\centering#2\strut}\hss
        \llap{\parbox[b]{\textwidth}{\raggedleft#3\strut}}%
    }
}

\newcommand{\reason}[2][\relax]{
  \ifthenelse{\equal{#1}{\relax}}{
    \left(\text{#2}\right)
  }{
    \left(\parbox{#1}{\raggedright #2}\right)
  }
}

\newcommand{\utag}[2]{\mathop{#2}\limits^{\text{(#1)}}}
\newcommand{\uref}[1]{(#1)}

\let\SavedDoubleVert\relax
\let\protect\relax
{\catcode`\|=\active
  \xdef\extendvert{\protect\expandafter\noexpand\csname extendvert \endcsname}
  \expandafter\gdef\csname extendvert \endcsname#1{\mskip-5mu \left.%
      \ifx\SavedDoubleVert\relax \let\SavedDoubleVert\|\fi
     \:{\let\|\SetDoubleVert
       \mathcode`\|32768\let|\SetVert
     #1}\:\right.\mskip-5mu}
}
\def\SetVert{\@ifnextchar|{\|\@gobble}
    {\egroup\;\mid@vertical\;\bgroup}}
\def\SetDoubleVert{\egroup\;\mid@dblvertical\;\bgroup}

\begingroup
 \edef\@tempa{\meaning\middle}
 \edef\@tempb{\string\middle}
\expandafter \endgroup \ifx\@tempa\@tempb
 \def\mid@vertical{\middle|}
 \def\mid@dblvertical{\middle\SavedDoubleVert}
\else
 \def\mid@vertical{\mskip1mu\vrule\mskip1mu}
 \def\mid@dblvertical{\mskip1mu\vrule\mskip2.5mu\vrule\mskip1mu}
\fi

\makeatother

\usepackage{fouridx}
\usepackage{framed}
\usetikzlibrary{positioning,matrix}

\usepackage{paralist}
\usepackage{enumerate}

\usepackage[normalem]{ulem}

{\endMakeFramed}

\newenvironment{ybox}{
	\setlength{\FrameSep}{1.5mm}
	\setlength{\FrameRule}{0mm}
  \MakeFramed {\FrameRestore}}%
{\endMakeFramed}

\newenvironment{gbox}{
	\setlength{\FrameSep}{1.5mm}
\setlength{\FrameRule}{0mm}
  \MakeFramed {\FrameRestore}}%
{\endMakeFramed}

{\endMakeFramed}

 {\endMakeFramed}

\usepackage{enumitem}

\usepackage{etoolbox}

\let\theparentequation\theequation
\patchcmd{\theparentequation}{equation}{parentequation}{}{}

\renewenvironment{subequations}[1][]{
	\refstepcounter{equation}%
	\setcounter{parentequation}{\value{equation}}
	\setcounter{equation}{0}
	\def\theequation{\theparentequation\alph{equation}}%
	\let\parentlabel\label
	\ifx\\#1\\\relax\else\label{#1}\fi
	\ignorespaces
}{%
	\setcounter{equation}{\value{parentequation}}
	\ignorespacesafterend
}

\newcommand*{\nextParentEquation}[1][]{
	\refstepcounter{parentequation}
	\setcounter{equation}{0}
	\ifx\\#1\\\relax\else\parentlabel{#1}\fi
}

\newcommand{\RCO}{R_{\op{CO}}}
\newcommand{\rCO}{r_{\op{CO}}}
\newcommand{\RS}{R_{\op{S}}}
\newcommand{\rS}{r_{\op{S}}}

\newcommand{\CS}{C_{\op{S}}}
\newcommand{\cS}{c_{\op{S}}}

\usepackage[outline]{contour}
\contourlength{1.2pt}

\title{One-Shot Perfect Secret Key Agreement\\ for Finite Linear Sources}

\author{Chung Chan, Navin Kashyap, Praneeth Kumar Vippathalla and Qiaoqiao Zhou
	\thanks{C.\ Chan (corresponding author, email: chung.chan@cityu.edu.hk) is with the Department of Computer Science, City University of Hong Kong.}
    \thanks{Q.\ Zhou (email: zq115@ie.cuhk.edu.hk) is with the Institute of Network Coding and the Department of Information Engineering, The Chinese University of Hong Kong.
	}
	\thanks{N.\ Kashyap (nkashyap@iisc.ac.in) and Praneeth Kumar V.\ (praneethv@iisc.ac.in) are with the Department of Electrical Communication Engineering, Indian Institute of Science, Bangalore 560012.}}

\begin{document}

\IEEEoverridecommandlockouts
\maketitle

\begin{abstract}
  We consider a non-asymptotic (one-shot) version of the multiterminal secret key agreement problem on a finite linear source model. In this model, the observation of each terminal is a linear function of an underlying random vector composed of finitely many i.i.d.\ uniform random variables. Restricting the public discussion to be a linear function of the terminals' observations, we obtain a characterization of the communication complexity (minimum number of symbols of public discussion) of generating a secret key of maximum length. 
 The minimum discussion is achieved by a non-interactive protocol in which each terminal first does a linear processing of its own private observations, following which the terminals all execute a discussion-optimal communication-for-omniscience protocol. The secret key is finally obtained as a linear function of the vector of all observations.
  

\end{abstract} 

\section{Introduction}
\label{sec:introduction}

The problem of secret key agreement via public discussion was first formulated for two terminals by Maurer \cite{maurer93} and Ahlswede and Csisz\'ar \cite{ahlswede93}, and subsequently extended to multiple terminals by Csisz\'ar and Narayan \cite{csiszar04}. In the set-up of this problem, the terminals involved must agree upon a secret key based on correlated observations from a source, using interactive public discussion. The key must be kept information-theoretically secure from an eavesdropper having access to the public discussion. The conventional setting allows unlimited public discussion, and the aim is to agree upon a secret key of largest possible length. The problem formulation is in fact asymptotic in nature: the terminals observe an infinite sequence of i.i.d.\ realizations of the correlated source random variables, and the asymptotic secret key rate (number of symbols of secret key generated per source realization) must be as large as possible. The largest possible asymptotic key rate, termed the secrecy capacity, is by now quite well understood \cite{csiszar04,chan10md}.

A more difficult problem is to determine the secrecy capacity under a constraint on the amount or rate of public discussion allowed. Specifically, when the (asymptotic) rate of public discussion is bounded above by $R$, the problem is to determine the maximum achievable secret key rate $\CS(R)$, which we term the rate-constrained secrecy capacity. This problem was considered in the case of two terminals by Tyagi \cite{tyagi13} and Liu et al.\ \cite{liu17}. The primary focus of Tyagi \cite{tyagi13} was on the related problem of characterizing what we will call the communication complexity $\RS$, which is the least discussion rate needed to achieve the (unconstrained) secrecy capacity; he left open the rate-constrained secrecy capacity problem. Liu et al.\ \cite{liu17} gave a characterization of the achievable region of key and discussion rate pairs using a notion of $XY$-concave envelopes that they develop. They used their methods to give a precise description of the ratio $\frac{\CS(R)}{R}$ in the regime of $R \to 0$. 

The multiterminal $\CS(R)$ and $\RS$ problems were considered in our prior works \cite{MKS16,chan17isit,chan18oo,chan18tit-arxiv}. Among our contributions there were some general outer bounds on the achievable rate region, and upper and lower bounds on $\RS$; in the special case of the hypergraphical source model, we derived tighter upper bounds on $\RS$ and the ratio $\frac{\CS(R)}{R}$ valid for all $R > 0$. In the important special case of the pairwise independent network (PIN) model (see e.g.\ \cite{nitinawarat10}), our bounds were good enough to precisely characterize $\RS$ and $\CS(R)$.

In this paper, we make further progress on these problems by focusing on the (multiterminal) finite linear source model~\cite{chan11itw}, which generalizes the hypergraphical source and PIN models. In the finite linear model, the observation of each terminal is a linear function of an underlying random vector composed of finitely many i.i.d.\ uniform random variables. Furthermore, we consider a non-asymptotic, \emph{single-shot} version of the secret key agreement problem as opposed to the asymptotic version in \cite{chan11itw}. In this version, the terminals observe only one realization of the source, and after some public discussion, must agree (with probability $1$) upon a secret key that is statistically independent of the public communication. Single-shot analogues of the $\RS$ and $\CS(R)$ problems can be formulated in this setting --- see Section~\ref{sec:problem}. We study these problems with a view towards extending the results obtained for the single-shot setting to the asymptotic model. 

Courtade and Halford \cite{courtade16} formulated and analyzed the single-shot secret key generation problem for hypergraphical sources. They made a key assumption to facilitate their analysis, namely, that the communication from each terminal is a linear function of its observations. Under this restriction, they effectively resolved the single-shot $\RS$ and $\CS(R)$ problems for hypergraphical sources. Note that linear discussion was also considered in \cite{chan11itw,chan11delay} for finite linear sources, but the objective was to achieve the unconstrained secrecy capacity of the asymptotic model perfectly with a finite block length, so as to avoid excessive delay in generating the secret key.
 
Taking inspiration from \cite{courtade16}, we too restrict the public discussion to be a linear function of the terminals' observations. Under the linear discussion model, for finite linear sources, we obtain a characterization (Corollary~\ref{Cor:RS} in Section~\ref{sec:results}) of the communication complexity of generating a secret key of maximum length. The minimum discussion is achieved by a non-interactive protocol in which each terminal first does a linear processing of its own private observations, following which the terminals all execute a (single-shot) discussion-optimal communication-for-omniscience protocol on their linearly processed observations. At the end of this, each terminal is able to recover the observations of all the other terminals (omniscience), and it then applies a linear function to the entire vector of observations to obtain a maximum-length secret key.

The rest of the paper is organized as follows. Section~\ref{sec:problem} contains the formal problem formulation, Section~\ref{sec:eg} presents an illustrative example, and Section~\ref{sec:results} contains statements of the main results, complete proofs of which can be found in the 
\ifPAGELIMIT
full version of this paper \cite{chan19isit-arxiv}.
\else
appendices.
\fi
The paper ends in Section~\ref{sec:extensions} with a discussion of the possible ways in which the results could be extended to settings beyond that of our problem formulation.


\section{Problem formulation}
\label{sec:problem}

We use the sans serif font $\RK$ to represent a random variable with distribution $P_{\RK}$ and taking values from a set $K$.
We use the boldface uppercase $\MM$ for matrices and boldface lowercase san serif font for random row vectors $\RMx:=\bM \Rx_1 & \dots & \Rx_{\ell(\RMx)} \eM $ where $\ell(\RMx)$ denotes the length of the vector. We assume all the entries take values from the same finite field $\bbF_q$ of order $q$. We take logarithm $\log$ to base $q$ and so all the information quantities are in the units of $\log q$~bits.
For a finite set $B$, we use $\RMy_B:=\bM \RMy_{i_1} & \dots & \RMy_{i_{\abs{B}}} \eM$ to denote a row vector obtained by concatenating the row vectors $\RMy_i$'s for some enumeration $i_1,\dots,i_{\abs{B}}$ of the set $B$. We use the notation
\begin{align*}
\RMx\in \Span{\RMy_B}\text{ or }\Span{\RMy_{i_1},\dots,\RMy_{i_{\abs{B}}}}  
\end{align*}
to mean that there exists a deterministic matrix $\MM$ such that $\RMx = \RMy_B \MM$.

As in \cite{csiszar04}, the multiterminal secret key agreement problem consists of a finite set $V = \{1,2,\ldots,m\}$ of $m\geq 2$ users who want to share a secret key after some public discussion that can be eavesdropped by a wiretapper. The one-shot perfect linear secret key agreement (SKA) scheme consists of the following phases.

\vspace{.5em}\noindent\textbf{One-shot private observation:} Each user $i\in V$ observes the component $\RMz_i$ of a given finite linear source $\RMz_V$ defined in \cite{chan10md} with the requirement that
\begin{align}
  \RMz_i \in \Span*{\RMx} \quad \forall i\in V \label{eq:fls}
\end{align}
for some uniformly random vector $\RMx$ over $\bbF_q$. $\RMx$ is referred to as the base of $\RMz_V$. In the special case when $\RMz_i$ is a subvector of $\RMx$, $\RMz_V$ is called the hypergraphical source~\cite{chan10md}, which is the source model considered in \cite{courtade16}. Unlike the model in \cite{csiszar04} and \cite{chan11itw} where each user observes $n$ i.i.d.\ samples of the source, we consider the one-shot model as in \cite{courtade16,MiloIT2016} where each user only observes one sample.

\vspace{.5em}\noindent\textbf{Private randomization:} Each user $i\in V$ privately generates a random vector $\RMu_i$ over $\bbF_q$ independent of the source $\RMz_V$, i.e.,
\begin{align}
  P_{\RMu_V|\RMz_V} = \prod_{i\in V} P_{\RMu_i}.\label{eq:u}
\end{align}
Note that there is no restriction on the length nor the distribution of $\RMu_i$, and so the requirement that it must be a vector over $\bbF_q$ does not lose generality. Note also that such a randomization was not explicitly considered in the formulations of \cite{csiszar04,chan11itw,courtade16}.

\vspace{.5em}\noindent\textbf{Linear public discussion:} Each user $i\in V$ publicly reveals the message
\begin{align}
  \RMf_i\in \Span*{\RMu_i,\RMz_i}.
\end{align}
Hence, everyone including the wiretapper observes $\RMf_V$. Unlike \cite{csiszar04}, the discussion above is non-interactive as interaction is unnecessary for linear discussion as explained in~\cite{chan10phd}.\footnote{Suppose the discussion is interactive, i.e., a message, say $\Rf$, revealed in public by some user~$i\in V$ is a linear function $\psi(\RMu_i,\RMz_i,\tilde\RMf)$ of the private observations of user~$i$ as well as all the previously discussed messages denoted by $\tilde\RMf$. By linearity, we can rewrite $\Rf$ as $\psi(\RMu_i,\RMz_i,\M0)+\psi(\M0,\M0,\tilde\RMf)$ where $\M0$ denotes an all-zero vector of an appropriate length. Note that, given $\tilde\RMf$, there is a bijection between $\Rf$ and $\Rf':=\psi(\RMu_i,\RMz_i,\M0)$, and so user~$i$ can reveal $\Rf'$ instead of $\Rf$ in public without loss of generality, since $\Rf$ can be recovered from $\Rf'$ and other discussion messages $\tilde\RMf$. As $\Rf'$ does not depend on $\tilde\RMf$, we can convert any interactive discussion to a non-interactive discussion by replacing every discussion message $\Rf$ by the corresponding $\Rf'$.}

\vspace{.5em}\noindent\textbf{Secret key agreement}
After the public discussion, each user $i\in V$ attempts to agree on a secret key $\RK$ satisfying
\begin{align}
  H(\RK|\RMu_i,\RMz_i,\RMf_V) &= 0 \quad \forall i\in V\label{eq:recover}\\
  \log \abs{K} - H(\RK|\RMf_V) &= 0 \label{eq:secrecy}
\end{align}
where \eqref{eq:recover} is the recoverability constraint that requires the secret key to be perfectly recoverable by every user and \eqref{eq:secrecy} is the secrecy constraint that requires the key to be uniformly random and perfectly independent of the entire public discussion. Note that we do not assume apriori that $\RK$ is a linear function of the private source, and so the key length $\log \abs{K}$ is not required to be an integer.\footnote{Nevertheless, it will follow from Theorem~\ref{thm:opt} that $\RK$ can be chosen to be a linear function of the private source without loss of optimality, and so the key length must be an integer.} 

The objective is to characterize the set of achievable key lengths and discussion lengths. In particular, a quantity of interest is the \emph{constrained secrecy capacity} defined as
\begin{align}
  \cS(r) := \cS(\RMz_V,r) := \max \Set{\log \abs{K}\mid \ell(\RMf_V)\leq r}, \label{eq:CSR}
\end{align}
where the maximization is over all possible secret key agreement schemes subject to a constraint on the total public discussion length, $r$. (Note that we omit the argument $\RMz_V$ if there is no ambiguity.) Characterizing the entire curve of $\cS(r)$ is difficult even in the case of linear discussion, but some points on the curve can be characterized, such as $\cS(0)$ considered in \cite{chan18isit}. As in \cite{csiszar04,chan11itw,courtade16}, we also consider the \emph{unconstrained secrecy capacity} defined as
\begin{align}
  \cS:=\cS(\RMz_V):=\max\Set{\cS(r)\mid r\geq 0},\label{eq:CS}
\end{align}
which is the secrecy capacity without the constraint on the discussion length. The smallest discussion length required to achieve $\cS$ is denoted by
\begin{align}
   \rS :=  \rS(\RMz_V):=\inf\Set*{r\geq 0 \mid \cS(\RMz_V,r)= \cS(\RMz_V)}\label{eq:RS}
\end{align}
and referred to as the \emph{communication complexity}. As in \cite{csiszar04,courtade16}, we will characterize $\cS$ and $ \rS$ using the closely related problem of communication for omniscience defined as follows. The problem under the one-shot model for hypergraphical and finite linear sources is proposed in \cite{milosavljevic11,MiloIT2016} and referred to as the \emph{cooperative data exchange}.

\vspace{0.5em}\noindent\textbf{Omniscience:} We say that the public discussion achieves omniscience of $\RMz_V$ if
\begin{align}
  H(\RMz_V|\RMu_i,\RMz_i,\RMf_V) = 0 \quad \forall i\in V.\label{eq:omni}
\end{align}
The smallest length of communication for omniscience is defined as
\begin{align}
  \rCO:=\rCO(\RMz) := \min \ell(\RMf_V),\label{eq:RCO}
\end{align}
where the minimization is over all public discussion schemes subject to \eqref{eq:omni} in place of \eqref{eq:secrecy} and \eqref{eq:recover}. In \cite{csiszar04}, the secret key agreement scheme that achieves the capacity is by first achieving omniscience of $\RMz_V$ and then extracting the secret key as a function of $\RMz_V$, implying that the rate of communication for omniscience is no smaller than the communication complexity. We say that $\cS$ can be achieved via omniscience of $\RMz_V$.

\section{Motivating Example}
\label{sec:eg}

We will use the following example to illustrate the problem formulation and motivate our main results. Consider $V=\Set{1,2,3,4}$ and a finite linear source $\RMz_V$ (see \eqref{eq:fls}) over the binary field $\bbF_2$ with a base $\RMx$ of length $\ell(\RMx)=4$ as follows:
  \begin{align}
    \begin{split}
      \RMz_1 &:= \bM \Rx_1 & \Rx_2 \oplus \Rx_3 \eM\\
      \RMz_2 &:= \bM \Rx_1 & \Rx_2 \oplus \Rx_4 \eM\\
      \RMz_3 &:= \bM \Rx_1 \oplus \Rx_2 & \Rx_3 \eM\\
      \RMz_4 &:= \bM \Rx_1 \oplus \Rx_3 \oplus \Rx_4 \eM.
    \end{split}
  \end{align}
  A feasible secret key agreement scheme is to choose
  \begin{align}
    \RK = \Rx_1, \RMf_1= \bM \Rx_2\oplus \Rx_3 \eM, \text{ and }\RMf_2 = \bM \Rx_2\oplus \Rx_4\eM,
  \end{align}
  but without any private randomizations $\RMu_V$ and discussions $\RMf_3$ and $\RMf_4$ by users~$3$ and $4$. The secret key $\RK$ is perfectly recoverable by every user, i.e., satisfying \eqref{eq:recover}, since users~$1$ and $2$ directly observes the key bit $\Rx_1$, which can also be computed by users~$3$ and $4$ using their private sources and public discussion as follows
  \begin{align*}
    \Rx_1 &= \bM \Rx_1\oplus \Rx_2 & \Rx_3 & \Rx_2\oplus \Rx_3 \eM \bM 1 \\ 1 \\ 1 \eM  \\[-2.5em]
          &\kern1em \underbrace{\kern5em}_{\RMz_3} \kern1em \underbrace{\kern3em}_{\RMf_1} \notag\\
          &= \bM \Rx_1\oplus \Rx_3\oplus \Rx_4 & \Rx_2\oplus \Rx_3 & \Rx_2\oplus \Rx_4 \eM \bM 1 \\ 1 \\ 1 \eM.\\[-2.5em]
          &\kern1em \underbrace{\kern5em}_{\RMz_4} \kern1em \underbrace{\kern3em}_{\RMf_1} \kern1em \underbrace{\kern3em}_{\RMf_2} \notag
  \end{align*}
  The secrecy constraint~\eqref{eq:secrecy} also holds because $\log\abs{K}=1=H(\RK|\RMf_V)$, which follows from the definition of the base $\RMx$ that $\Rx_1$ is uniformly random and independent of $\Rx_2$, $\Rx_3$, and $\Rx_4$.

  Note that the above scheme does not achieve the omniscience condition in \eqref{eq:omni} because users~$1$, $2$ and $4$ cannot recover $\Rx_3$ after the discussion. However, it is easy to show that omniscience can be achieved if we further set $\RMf_3=\bM \Rx_3\eM$, i.e., with an additional bit of discussion by user~$3$. Since $1$ bit of secret key can be achieved with $2$ bits of public discussion and omniscience can be further achieved with an additional bit of discussion, we have
  \begin{align}
     \cS( \rCO) \geq
    \begin{cases}
      1 & r\geq 2\\
      0 & r< 2
    \end{cases}, \text{ and  }
          `1\{\begin{aligned}
             \cS &\geq 1\\
             \rS &\leq 2\\
             \rCO & \leq 3
          \end{aligned}`2.
  \end{align}
  by the definitions \eqref{eq:CSR}, \eqref{eq:CS}, \eqref{eq:RS} and \eqref{eq:RCO}. The challenge is to show the reverse inequalities and therefore establish the optimality of the achieving schemes.

\section{Main results}
\label{sec:results}

We start with some rather general admissible conditions that simplify the secret key agreement scheme significantly without loss of optimality.

\begin{Theorem}
  \label{thm:opt}
  $\cS(r)$ remains unchanged even if we set
  \begin{subequations}
  \label{eq:opt}
  \begin{align}
    \ell(\RMu_i) &= 0 \quad \forall i\in V, \quad\text{and}\label{eq:opt:u}\\
    \RK &= \RMk \in \Span*{\RMz_V},\label{eq:opt:k}
  \end{align}
  \end{subequations}
  which mean respectively that private randomizations are not needed and that the secret key can be chosen to be linear function of the private source.
\end{Theorem}

\begin{Corollary}
  \label{cor:opt}
  $\cS(r)$ must be integer, non-decreasing and right continuous in $r$.
\end{Corollary}

\begin{Proof}
  For the corollary, the fact that $\cS(r)$ must be an integer follows from \eqref{eq:opt:k} that the key can be linear and therefore a uniformly random vector by the secrecy constraint~\eqref{eq:secrecy}. Monontonicity and continuity follows directly from the definition~\eqref{eq:CSR}. The proof of the theorem is more involved and given in
  \ifPAGELIMIT
  \cite[Appendix~A]{chan19isit-arxiv}.
  \else 
  Appendix~\ref{proof:opt}.
  \fi 
\end{Proof}

For instance, the example in Section~\ref{sec:eg} considers such a secret key agreement scheme without private randomization. The secret key $\Rx_1$ is also linear in the private source trivially because it is directly observed by users~$1$ and $2$.
Note that our formulation allows the private randomizations to have arbitrary length and distribution, and the key to be arbitrary random variables that need not be linear in the private source.
The above admissible constraints~\eqref{eq:opt} makes the problem tractable as it significantly reduces the space of secret key agreement schemes we need to consider to characterize $\cS(r)$. Indeed, since there is only a finite number of linear functions of $\RMz_V$, there is only a finite number of admissible secret key agreement scheme. It is worth noting that the constraints~\eqref{eq:opt} were assumed in the formulation of \cite{courtade16} for the hypergraphical source model, and our result implies that such constraints are admissible since hypergraphical sources are special case of the finite linear sources. 

For the general source model in \cite{csiszar04}, the admissible constraint~\eqref{eq:opt:u} that private randomization does not help improve $\cS(r)$ remains a plausible conjecture. However, it is clear that the constraint \eqref{eq:opt:k} is not admissible for some sources that are not finite linear. Nevertheless, this constraint is essential in bringing the existing characterizations of the capacity from the general source model to the one-shot finite linear source model as follows.

\begin{Theorem}
  \label{thm:CSRCO}
  $\cS(r)$ in the extreme cases with $0$ and respectively unbounded discussion lengths are
  \begin{align}
    \cS(0) &= \max\Set{H(\RMg)\mid \RMg\in \Span{\RMz_i},\forall i\in V} \label{eq:CS=JGK}\\
    \cS &= \floor*{\min_{\mcP\in \Pi'(V)} \frac{\sum_{C\in \mcP}
    H(\RMz_C)-H(\RMz_V)}{\abs{\mcP}-1}},\label{eq:CS=I}
  \end{align}
  where the maximization is over the choices of random vector $\RMg$, and the minimization is over the collection $\Pi'(V)$ of partitions of $V$ into at least two non-empty disjoint sets.
  Furthermore, $\cS$ can be achieved via communication for omniscience of $\RMz_V$ at the smallest length
  \begin{align}
    \rCO &= H(\RMz_V) - \cS,\label{eq:CSRCO}
  \end{align}
  which implies the upper bound $\rS\leq \rCO$ on $\rS$.
\end{Theorem}

\begin{Proof}
  \ifPAGELIMIT
  See \cite[Appendix~B]{chan19isit-arxiv}.
  \else 
  See Appendix~\ref{proof:CSRCO}.
  \fi 
\end{Proof}

For the running example given in Section~\ref{sec:eg}, it is straightforward to evaluate the above expressions \eqref{eq:CS=JGK}, \eqref{eq:CS=I} and \eqref{eq:CSRCO} to yield $\cS(0)=0$, $\cS=1$ and $\rCO=3$. In particular, an optimal solution to \eqref{eq:CS=I} can be shown to be $\mcP=\Set{\Set{1,2,3},\Set{4}}$. This implies the optimality of the omniscience scheme in Section~\ref{sec:eg} in achieving both $\cS$ and $\rCO$. 

The above result follows quite directly from existing works for the asymptotic model. For instance, the r.h.s.\ of \eqref{eq:CS=JGK} is the multivariate G\'acs-K\"orner common information
evaluated for the finite linear source model. $\RMg$ is called the maximum common function of $\RMz_i$ for $i\in V$. 
$\cS(0)=J_{\op{GK}}(\RMz_V)$ was shown in \cite{chan18isit} but for the asymptotic model instead. It is straightforward to extend this result to the current one-shot model.

The duality~\eqref{eq:CSRCO} between secret key agreement and communication for omniscience also follows directly from the asymptotic model in \cite{csiszar04}, which is specialized to the asymptotic finite linear source model in \cite{chan11itw}. The characterization \eqref{eq:CS=I} of $\cS$ is the same as that of the asymptotic model~\cite{csiszar04,chan10md} except for the floor operation, since the minimization in \eqref{eq:CS=I} may not be integer but $\cS$ must be integer by Corollary~\ref{cor:opt}. The characterization of $\rCO$ for the one-shot finite linear source model is given in \cite{MiloIT2016,ding18it}, which focus primarily on the omniscience problem instead of the secret key agreement problem.

Note that one can summarize the theorem by saying that $\cS(0)$, $\cS$ and $\rCO$ for the one-shot model is the same as those of the asymptotic model for finite linear source but with an additional integer constraint: $\CS(0)$ is already an integer for the asymptotic model while we can take the floor and the ceiling respectively for $\CS$ and $\RCO$ to turn them into integer achievable lengths. It therefore appears reasonable to conjecture that $\cS(r)$ for the one-shot model is the same as the $\CS(R)$ for the asymptotic model for finite linear source but with an additional floor operation as in \eqref{eq:CS=I} to satisfy the integer constraint in Corollary~\ref{cor:opt}. The following result resolves this partially at the communication complexity $\rS$.

\begin{Theorem}
  \label{thm:RS}
  If $\rS<\rCO$, then there exists $\RMz'_V$ with
  \begin{align}
    \RMz'_i &\in \Span*{\RMz_i}\quad \forall i\in V
  \end{align}
  such that
  \begin{align}
    \rS(\RMz'_V) &= \rS(\RMz_V)\\
    \rCO(\RMz'_V) &< \rCO(\RMz_V)\\
    \cS(\RMz'_V) &= \cS(\RMz_V).
  \end{align}
  $\RMz'_V$ is said to be reduced source of $\RMz_V$ (by linear processing), since the above implies $H(\RMz'_V)<H(\RMz_V)$.
\end{Theorem}

\begin{Corollary}
  \label{Cor:RS}
  The communication complexity is
  \begin{align}
    \rS = \min\Set*{\rCO(\RMz'_V)\mid \RMz'_i\in \Span{\RMz_i},\cS(\RMz'_V)=\cS(\RMz_V)},\label{eq:RSRCO}
  \end{align}
  achieved via omniscience of the linearly reduced source $\RMz'_V$.
\end{Corollary}

\begin{Proof}
  The corollary follows immediately from theorem by repeatedly linearly reducing the source until $\rS=\rCO$. This is possible since the theorem guarantees linear processing of the source exists that can reduce $\rCO$ without changing $(\cS,\rS)$ whenever $\rS<\rCO$.
  For the proof of the theorem, 
  \ifPAGELIMIT
  see \cite[Appendix~C]{chan19isit-arxiv}.
  \else 
  see Appendix~\ref{proof:RS}.
  \fi 
\end{Proof}

For the running example in Section~\ref{sec:eg}, the omniscience scheme does not achieve $\rS$ as $\rS\leq 2$ by the secret key agreement scheme without omniscience described in Section~\ref{sec:eg}. As $\rS<\rCO$, the theorem above guarantees a linear processing of the source that reduces $\rCO$ without changing $(\cS,\rS)$. Such a linearly reduced source can be obtained with 
\begin{align*}
\RMz'_3=\RMz_3 \bM 1 \\ 1 \eM= \Rx_1\oplus \Rx_2\oplus \Rx_3 \in \Span*{\RMz_3}   
\end{align*}
and $\RMz'_i=\RMz_i$ for $i\in {1,2,4}$. It is straightforward to show that $\cS(\RMz'_V)=1$ by \eqref{eq:CS=I} and $\rCO(\RMz'_V)=2$ by \eqref{eq:CSRCO}. Note the source is reduced in the sense that $H(\RMz'_V)=3<H(\RMz_V)$. By going through all possible independent linear processings of $\RMz_i$'s, which is possible as there is only a finite number of possibilities, one can show that the above defined $\RMz'_V$ is optimal to \eqref{eq:RSRCO} achieving the minimum $\rCO$, and so $\rS=2$ as desired by the above corollary.

Note that the characterization of $\RS$ remains open for the asymptotic model~\cite{csiszar04} but we believe that it can be resolved for finite linear source model by extending the above result to the asymptotic case. This means in particular that $\rS$ for the one-shot model is the ceiling of the $\RS$ for the asymptotic model. In Section~\ref{sec:extensions}, we outline the challenges involved in such extension.

\section{Extensions}
\label{sec:extensions}


In this work, we considered the one-shot secret key agreement problem under a finite linear source model with linear public discussion, perfect secrecy and recoverability. However, we believe that all the results can be extended without assuming the discussion is linear. In particular, extending Theorem~\ref{thm:CSRCO} is straightforward as the converse parts follow from those of the asymptotic model without requiring the discussion to be linear. Extending Theorem~\ref{thm:opt} and Theorem~\ref{thm:RS} appears challenging. The current proofs rely on the linearity of discussion.

Another possible extension of the current results is to the asymptotic model where users observe $n\geq 1$ i.i.d.\ samples $\RMz_V^n:=\bM \RMz_{V1} &\dots&\RMz_{Vn}\eM$ of the private source, and the constrained secrecy capacity and discussion rate is per sample of the observation, i.e.,
\begin{align*}
    \CS(R) := \max\Set*{`1.\frac{\log\abs{K}}n`2| \frac{\ell(\RMf_V)}n \leq R}.
\end{align*}
The recoverability~\eqref{eq:recover} and secrecy~\eqref{eq:secrecy} constraints can also be relaxed to the asymptotic versions in \cite{csiszar04}, i.e., for some positive $`d_n,`e_n\to 0$ as $n\to `8$,
\begin{align*}
    &\frac1n\log\abs{K}-H(\RK|\RMf_V) \leq `d_n\\
    &\Pr`1(\exists i\in V, \RK\neq `f_i(\RMu,\RMz_i^n,\RMf_V)`2)\leq `e_n
\end{align*}
for a sequence in $n$ of secret key agreement schemes and some functions $`f_i$ for $i\in V$ that user $i$ uses to recover the secret key. As mentioned below Theorem~\ref{thm:CSRCO}, the characterizations of $\CS(0)$, $\CS$ and $\RCO$ are already known for the asymptotic model and they are indeed used to derive the corresponding characterizations for the one-shot model. 
We believe that the other results in Theorem~\ref{thm:opt} and Theorem~\ref{thm:RS} can be extended. The current proofs can be directly extended if we impose perfect recoverability instead, i.e., with $`e_n=0$ for sufficiently large $n$. However, the proofs without assuming perfect recoverability remain elusive. What we desire is a proof that perfect recoverability is admissible and can therefore be be assumed without loss of optimality. In the similar vein, we also desire a proof that $\RS$ can be achieved exactly, i.e., for sufficiently large $n$, there exists a secret key agreement scheme with $\frac1n \log\abs{K}= \CS$ and $\frac{\ell(\RMf_V)}{n}=\RS$, and that linear public discussion is admissible.

\bibliographystyle{IEEEtran}
\bibliography{IEEEabrv,ref}

\appendix

\section{Proofs}
\label{sec:proofs}

\ifPAGELIMIT
Give proof sketches here.
\else 
\subsection{Proof of Theorem~\ref{thm:opt}}
\label{proof:opt}

First we will show the admission constraint \eqref{eq:opt:k}, i.e., $\cS(r)$ remains unchanged when the secret key is chosen to be linear function of the private source. Consider any optimal SKA scheme with a fixed discussion length $r$, i.e., with secret key $\RK$ having length $\cS(r)$ and discussion $\RMf_V$ having length $r$. Define 
\[\RMy_i := \bM \RMu_i & \RMz_{i} & \RMf_V \eM.\]
By the recoverability constraint \eqref{eq:recover} of the secret key, $\RK$ is a common function of $\RMy_i$ for $i\in V$. Trivially, $\RMf_V$ is also a common function of $\RMy_i$. Let $\RMg$ be a maximum common function, as defined for Theorem \ref{thm:CSRCO}, of $\RMy_i$'s instead of $\RMz_i$'s. From \cite{gacs72}, we know that any common function of $\RMy_i$ is a function of $\RMg$. Hence $(\RK,\RMf_V)$ is a function of $\RMg$ i.e., 
 \begin{align}
    H(\RK,\RMf_V|\RMg) = 0\label{eq:lin:cmf}.
 \end{align} 
 It was shown in \cite{chan18isit} that $\RMg$ is a linear function for a finite linear source, i.e., $\RMg \in \Span*{\RMy_i} \text{ or } \Span*{\RMu_i,\RMz_i,\RMf_V}, \forall i\in V$. Therefore,
 \begin{align*}
    \RMg &\in \Span*{\RMu_V,\RMz_V}.
  \end{align*}
  In fact, $\RMf_V$ is a linear function of $\RMg$ because of the linearity of the communication, $\RMf_V \in \Span*{\RMu_V,\RMz_V}$, and the linearity of the maximal common function $\RMg$ . We can therefore write, $\RMf_V \in \Span*{\RMg}$.
  
  We will show that the secret key rate remains unchanged if we choose the secret key to be the linear function $\RK'=\RMk' \in \Span*{\RMg}$ such that
   \begin{align}
    \RMg &\in \Span*{\RMf_V,\RMk'},\label{eq:lin:gk}\\
    \log|K'|&= H(\RK'|\RMf_V)\label{eq:lin:key}.
  \end{align}
  Note that the above choice of $\RK'$, if exists, is a feasible choice of secret key because \eqref{eq:lin:key} implies the secrecy constraint~\eqref{eq:secrecy} while the recoverability constraint~\eqref{eq:recover} follows from the fact that $\RMk'$ is a function of the maximum common function $\RMg$. 
  
  To show that $\RK'$ exists, consider $\RMf_V=\RMg \MW$ for some matrix $\MW$ and set $\RMk'=\RMg\MN$ where $\MN$ is a matrix whose column space is the left null space of $\MW$. \eqref{eq:lin:gk} then follows from the fact that $\RMg \bM \MW & \MN\eM = \bM \RMf_V & \RMk'\eM$ is a bijection as $\bM \MW & \MN\eM$ has full column rank.
  To show \eqref{eq:lin:key}, note that the columns of $\MN$ cannot be spanned by the columns of $\MW$, and so $\RMk'$ is independent of $\RMf_V$, i.e., \[H(\RMk')=H(\RMk'|\RMf_V)=H(\RK'|\RMf_V).\]
  It remains to show that $\RMk'$ is uniformly distributed, i.e.,
  \[\log\abs{K'}=\ell(\RMk')=H(\RMk').\]
  Notice we can choose $\MN$ to have full column rank, in which case $\RMk'=\RMg\MN$ is uniformly distributed if $\RMg$ is. Since $\RMy_V\in \Span{\RMx}$, we can write $\RMg$ as $\RMg = \RMx\MM$ for some matrix $\MM$. We can choose $\MM$ to have full column rank, which then implies that $\RMg$ is uniformly distributed as desired because $\RMx$ is.

  Finally, we argue as follows that the secret key rate is not diminished if $\RK'$ is used as the secret key instead.
    \begin{align}
       \log|K'|\utag{a}= H(\RK'|\RMf_V)\utag{b}\geq H(\RMg|\RMf_V)\utag{c}\geq H(\RK|\RMf_V)\utag{d}=\log|K|,
    \end{align}
  where \uref{a} is from \eqref{eq:lin:key}, \uref{b} is due to the fact  that $\RMg$ is a linear function of $\RK'$ and $\RMf_V$ \eqref{eq:lin:gk}, \uref{c} is from \eqref{eq:lin:cmf} and \uref{d} follows from the perfect secrecy \eqref{eq:secrecy} of $\RK$. 
  
Next, we impose the linearity~\eqref{eq:opt:k} of the key and show that that the other constraint~\eqref{eq:opt:u} is admissible, i.e., private randomization is not needed. Consider any user $j \in V$, and rewrite $\bM \RMu_j&\RMz_j\eM$ as $\bM \Ru & \RMv \eM$ by reordering the components such that $\Ru$ is a component of  $\RMu_j$ and $\RMv$ contains the rest of the components of $\RMu_j$ and $\RMz_j$. We will argue that $\Ru$ can be removed without affecting the secret key rate or the discussion rate. Consider the case where $\RMf_j$ does not depend on $\Ru$. Then, $\RMk$ also cannot depend on $\Ru$, or the recoverability constraint~\eqref{eq:recover} fails. Hence $\Ru$ can be removed as desired. Consider the non-trivial case where there is a component $\Rw$ of $\RMf_j$ such that 
\[\Rw = \alpha.\Ru - \beta(\RMv)\]
for some $\alpha \in \mathbb{F}_q\backslash\{0\}$ and linear function $\beta$. We can also write 
    \begin{align*}
    \RMk&= \alpha'(\Ru) + \beta'(\RMv)+\gamma'(\RMf_{V`/\Set{j}}),\\
    \RMf_j&= \alpha''(\Ru) + \beta''(\RMv),
  \end{align*}
   for some linear functions $\alpha',\alpha'',\beta',\beta''\text{ and } \gamma'$. Define for $i \in V$,
     \begin{align*}
    \RMf'_i&:= 
    \begin{cases}
    \RMf_i-\alpha''\left(\frac{\Rw}{\alpha}\right) = \alpha''\left(\frac{\beta(\RMv)}{\alpha}\right) + \beta''(\RMv), & \text{if $i=j$},\\
    \RMf_i, & \text{otherwise},
  \end{cases}\\
  \RMk' &:= \RMk-\alpha'\left(\frac{\Rw}{\alpha}\right) = \alpha'\left(\frac{\beta(\RMv)}{\alpha}\right) + \beta'(\RMv)+\gamma'(\RMf'_{V`/\Set{j}}).
  \end{align*}
  Note that both $\RMk'$ and $\RMf'_V$ are independent of $\Ru$. Furthermore, 
  \begin{align*}
   \ell(\RMk) &=\ell(\RMk')\\
   \ell(\RMf_V) &=\ell(\RMf'_V)\\
   H(\RMk|\RMf_V) &\utag{a}=H\left(`1.\RMk-\alpha'\left(\frac{\Rw}{\alpha}\right)`2|\RMf_V\right)\\
   &\utag{b}=H(\RMk'|\RMf'_V)
  \end{align*}
  where \uref{a} follows from the fact that $\Rw$ is a component of $\RMf_i$ and \uref{b} is because $H(\RMf'_i|\RMf_i)=0$ for all $ i \in V$.
  Hence $(\RMk', \RMf'_V)$ is an optimal SKA scheme which  does not depend on $\Ru$, i.e., $\Ru$ can be removed.

\subsection{Proof of Theorem~\ref{thm:CSRCO}}
\label{proof:CSRCO}

With no public discussion, i.e., $\RMf_V=\emptyset$, by the recoverability constrain~\eqref{eq:recover}, the secret key must be a function of the G\'acs--K\"orner common information~\cite{gacs72} of $\RMz_V$. For finite linear source, \cite[Theorem~4.2]{chan18isit} showed that the latter quantity equals to the r.h.s. of~\eqref{eq:CS=JGK}, thereby establishing $\leq$ for~\eqref{eq:CS=JGK}. To prove the reverse inequality, note that, we can write the solution to \eqref{eq:CS=JGK} as $\RMg=\Mx\MM$ for some matrix $\MM$ because $\RMz_V\in \Span{\RMx}$. Furthermore, $\MM$ can be chosen to have have full column rank without loss of optimality, and so $\RMg$ can be uniformly distributed.  Then, the desired secrecy capacity can be achieved with $\RK=\RMg$, where the uniformity of $\RMg$ implies the secrecy constraint~\eqref{eq:secrecy} and the recoverability constraint~\eqref{eq:recover} also follows trivially from the fact that $\RMg$ is a common function. 

\emph{Converse Part.} By Thorem~\ref{thm:opt}, we can assume $\ell(\RMu_i) = 0, \forall i\in V$ and $\RK = \RMk \in \Span*{\RMz_V}$ without loss of optimality. Now, since $\RMk$ and $\RMf_V$ are functions of $\RMz_V$,
\begin{align*}
H(\RMz_V)&=H(\RMf_V,\RMk,\RMz_V)\\
		 &=H(\RMf_V)+H(\RMk|\RMf_V)+H(\RMz_V|\RMf_V,\RMk)\\
		 &=\sum_{i=1}^{m}H(\RMf_i|\RMf_{[i-1]})+H(\RMk|\RMf_V)+\sum_{i=1}^{m}H(\RMz_i|\RMf_V,\RMk,\RMz_{[i-1]})
\end{align*}
Setting 
\begin{align*}
r_i=H(\RMf_i|\RMf_{[i-1]})+H(\RMz_i|\RMf_V,\RMk,\RMz_{[i-1]}),
\end{align*} 
the previous equality yields 
\begin{align*}
H(\RMk|\RMf_V)=H(\RMz_V)-\sum_{i=1}^{m}r_i.
\end{align*} 
Next, we show that $r_V=(r_1,\dots,r_m)\in\rsfsR(\RMz_V)$, where
\begin{align*}
\rsfsR(\RMz_V):=\Set{r_V\mid r(B)\geq H(\RMz_B|\RMz_{V\setminus B}),\forall B\subsetneq V}.
\end{align*} 
To that end, 
\begin{align*}
 H(\RMz_B|\RMz_{V\setminus B})&=H(\RMf_V,\RMk,\RMz_B|\RMz_{V\setminus B})\\
 						   &=H(\RMf_V|\RMz_{V\setminus B})+H(\RMk|\RMz_{V\setminus B},\RMf_V)\\
						   &\kern1em+H(\RMz_B|\RMz_{V\setminus B},\RMf_V,\RMk,)\\
						   &=\sum_{i=1}^{m}H(\RMf_i|\RMz_{V\setminus B},\RMf_{[i-1]})+H(\RMk|\RMz_{V\setminus B},\RMf_V)\\
						   &\kern1em+\sum_{i\in B}H(\RMz_i|\RMz_{V\setminus B\cup[i-1]},\RMf_V,\RMk)\\
						   &\leq \sum_{i\in B}H(\RMf_i|\RMf_{[i-1]})+\sum_{i\in B}H(\RMz_i|\RMz_{[i-1]},\RMf_V,\RMk)\\
						   &=r(B)
 \end{align*} 
where the last inequality is because 
\begin{align*}
H(\RMf_i|\RMz_{V\setminus B},\RMf_{[i-1]}) 
\begin{cases}	
=0 &i\not\in B,	\\
\leq H(\RMf_i|\RMf_{[i-1]}) &i\in B,
\end{cases}
 \end{align*} 
and $H(\RMk|\RMz_{V\setminus B},\RMf_V)=0$ by perfect recoverability. 
Now, since $H(\RMk)$ is an integer, for $r_V\in\rsfsR(\RMz_V)$, 
\begin{align*}
H(\RMk)=H(\RMk|\RMf_V)&=\floor*{H(\RMz_V)-\sum_{i=1}^{m}r_i}\\	
			 &\leq \floor*{H(\RMz_V)-\min_{r_V\in\rsfsR(\RMz_V)}r(V)}\\
			 &=\floor*{\min_{\mcP\in \Pi'(V)} \frac{\sum_{C\in \mcP} H(\RMz_C)-H(\RMz_V)}{\abs{\mcP}-1}}
\end{align*} 

\emph{Achievability Part.} 
For any vector $r_V\in \rsfsR(\RMz_V)\cap \mathbb{Z}^{m}$, it was shown by~\cite[Theorem~1]{MiloIT2016}  that there exists a corresponding linear noninteractive discussion scheme which renders omniscience. 
Further, it was shown by~\cite[Corollary~6]{ding18it} that 
\begin{align}
\label{eq:ILP:LP:gap<1}
\min_{r_V\in\rsfsR(\RMz_V)\cap\mathbb{Z}^{m}}r(V)
=\ceil*{\max_{\mcP\in \Pi'(V)}\sum_{C\in \mcP}\frac{H(\RMz_V)-H(\RMz_C)}{\abs{\mcP}-1}}.
\end{align}
Consider any optimal solution $r_V^*$ to the l.h.s. of~\eqref{eq:ILP:LP:gap<1}. Denote the corresponding linear noninteractive discuss that attains omniscience by $\RMf_V=\RMz_V\ML=\RMx\bM \MM_1 & \MM_2 & \cdots & \MM_m \eM\ML=\RMx\MT$. 

Now, it remains to extract a perfect secret key from the omniscience obtained above. For each realization $\Mf_V$ of $\RMf_V$, let $\mathcal{P}_{\Mf_V}=\Set{\Mx\mid \Mx\MT=\Mf_V}$ be the set of all $\Mx$ which generate $\Mf_V$.
By the definition of the finite linear source, observe that each entry $\Rx_i$ of $\RMx$ is i.i.d. uniform from $\mathbb{F}_q$, i.e., $\Pr(\RMx=\Mx)=q^{-\ell(\RMx)},\forall \Mx$. By the linearity of the discussion above, it is easy to see that 
$|\mathcal{P}_{\Mf_V}|$ is the same for all realizations of $\RMf_V$. Since $\RMf_V$ has dimension $r^*(V)$, $|\mathcal{P}_{\Mf_V}|=q^{\ell(\RMx)-r^*(V)},\forall \Mf_V$. Set $K=\mathbb{F}_q^{\ell(\RMx)-r^*(V)}$. For each $\Mf_V$, label each $\Mx\in\mathcal{P}_{\Mf_V}$ with a unique element in $K$. Then, upon observing $\Mf_V$, every user which knows $\Mx$ by omniscience picks the label of $\Mx$ as the secret key. Since $\RMx$ is uniformly distributed and $|\mathcal{P}_{\Mf_V}|$ has the same size,  it follows that this random label is uniformly distributed over $K$ and independent of $\RMf_V$, thereby constituting a perfect secret key. Therefore, 
\begin{align*}
\cS &\geq \ell(\RMx)-r^*(V)\\
       &=H(\RMz)-\ceil*{\max_{\mcP\in \Pi'(V)}\sum_{C\in \mcP}\frac{H(\RMz_V)-H(\RMz_C)}{\abs{\mcP}-1}}\\
       &=\floor*{\min_{\mcP\in \Pi'(V)} \frac{\sum_{C\in \mcP} H(\RMz_C)-H(\RMz_V)}{\abs{\mcP}-1}}.
\end{align*} 
This completes the proof.

\subsection{Proof of Theorem~\ref{thm:RS}}
\label{proof:RS}

To prove this theorem, we will need the following lemma. We say that $\RMz_V'$ can be \emph{simulated} by the source $\RMz_V$ if $\RMz'_i \in \Span*{\RMz_i} \text{ for all } i\in V$.

\begin{Lemma}
\label{lem:sim}
If $\RMz_V'$ can be simulated by $\RMz_V$, then  $\rS(\RMz'_V) \geq \rS(\RMz_V)$.  
\end{Lemma}
\begin{proof}
Any secret key generation protocol for $\RMz'_V$ can be simulated by $\RMz_V$.
\end{proof}

Now consider the finite linear source $\RMz_V = \bM \RMz_1&\ldots&\RMz_m\eM$ on the  set of users $V$, defined by $\RMz_i \in \Span*{\RMx}$ or equivalently, $ \RMz_i = \RMx \MM_i $ for each $i \in V$, where $\MM_i$ is a $\ell \times t_i$ matrix over $\bbF_q$. Set $\MM := \bM \MM_1 & \MM_2 & \cdots & \MM_m \eM$, so that 
\begin{align*}
    \RMz_V &= \RMx \MM.
\end{align*}
We assume, without loss of generality, that $\MM$ has full row rank, so that $H(\RMz_V) = H(\RMx) = \ell$. We will let $Z_i$ (resp.\ $Z_V$) denote the row-space of $\MM_i$ (resp.\ $\MM$). Since $\text{rank}_{\bbF_q}(\MM) = \ell$, we have $\dim_{\bbF_q}(Z_V) = \ell$, so that $Z_V \cong \bbF_q^\ell$.

Let $\RMf_V$ be a (non-interactive) linear communication protocol, i.e., $\RMf_V \in \Span*{\RMz_V}$, that generates a secret key $\RK$ using $\rS(\RMz_V)$ symbols from $\bbF_q$ as public communication. Since $\rS(\RMz_V)<\rCO(\RMz_V)$, the communication $\RMf_V$ is insufficient for omniscience. Assume that user $1$ cannot recover all of $\RMz_V$ from $(\RMz_1 , \RMf_V)$. Then, via linearity of the source and the communication, there exists an observation 
\begin{align*}
\Mz_V=\bM \Mz_1 &\Mz_2 &\ldots&\Mz_m \eM \in Z_V,\quad \Mz_V \ne \pmb{0},
\end{align*}
such that $\bM \Mz_1 & \Mz_V \MA \eM = \pmb{0}$. 

Fix a basis $\Mb_1,\Mb_2,\ldots,\Mb_\ell$ for $Z_V$, with $\Mb_{\ell} = \Mz_V$ as above. Let $\widetilde{\MM}$ be the matrix having $\Mb_1,\Mb_2,\ldots,\Mb_{\ell}$, in that order, as its $\ell$ rows. Thus, $\text{row-space}(\widetilde{\MM}) = \text{row-space}(\MM) = Z_V$. There is then an invertible (change-of-basis) matrix $\MB$ such that $\MM = \MB\widetilde{\MM}$. We can now write 
\begin{align*}
\RMz_V & = \RMx \MM = \RMx \MB \widetilde{\MM} = \RMy \widetilde{\MM},
\end{align*}
where $\RMy := \RMx\MB$. Since $\Rx_1,\ldots,\Rx_\ell$ are i.i.d. $\text{Unif}(\bbF_q)$ rvs, and $\MB$ is invertible, $\RMy = \bM \Ry_1&\Ry_2&\ldots&\Ry_\ell\eM$ is also composed of i.i.d. rvs $\Ry_i$ uniformly distributed over $\bbF_q$. Note that the relation $\RMy = \RMx \MB$ allows each $\Ry_j$ to be expressed as a linear combination of the $\Rx_i$s, and vice versa. Thus,
\begin{align*}
\RMz_V & = \sum_{i=1}^\ell \Ry_i \Mb_i,
\end{align*}
provides an alternative description of the finite linear source $\RMz_V$.

Since $\Mb_\ell = \Mz_V$, we have 
\begin{align*}
    \RMz_V & = \sum_{i=1}^{\ell-1} \Ry_i \Mb_i + \Ry_\ell \Mz_V.
\end{align*}
Since $\Mz_1 = \pmb{0}$ (by choice of $\Mz_V$), we see that $\RMz_1$ consists of linear combinations of $\Ry_1,\ldots,\Ry_{\ell-1}$ alone. Also, since $\Mz_V \MA = \pmb{0}$, the communication 
\begin{align*}
    \RMf_V &= \RMz_V \MA = \bigl(\sum_{i=1}^{\ell-1} \Ry_i \Mb_i + \Ry_\ell \Mz_V\bigr) \MA = \sum_{i=1}^{\ell-1} \Ry_i \Mb_i \MA
\end{align*} consists of linear combinations of $\Ry_1,\ldots,\Ry_{\ell-1}$ alone. Consequently, the secret key $\RK$, which must be generated by user $1$ from $(\RMz_1,\RMf_V)$, is a function of $\bM \Ry_1&\ldots&\Ry_{\ell-1}\eM$ alone.

Now, for each $i \in V$, we have 
\begin{align*}
    \RMz_i & = \RMx \MM_i = \RMy \widetilde{\MM}_i,
\end{align*}
with $\widetilde{\MM}_i :=\MB^{-1}\MM_i$ again being a $\ell \times t_i$ matrix. Define 
\begin{align*}
    V_0 & := \{i \in V: \text{ the last row of } \widetilde{\MM}_i \text{ is zero}\}.
\end{align*}
For each $i \in V_0$, $\RMz_i$ is a function of $\bM \Ry_1&\ldots&\Ry_{\ell -1}\eM$ alone; in this case, set $\MM'_i = \widetilde{\MM}_i$. On the other hand, for each $i \notin V_0$, use elementary column operations to convert $\widetilde{\MM}_i$ into an $\ell \times t_i$ matrix $\MM'_i$, in which the last row has a single nonzero entry, occurring in the last column: 
\begin{align*}
    \MM'_i[\ell,j] & =
    \begin{cases}
    0 ,  \quad j = 1,\ldots,t_i-1 = 1,\\
    1 ,  \quad j = l.
    \end{cases}
\end{align*}
Note that for each $i \in V$, we can express$\MM'_i$ as $\widetilde{\MM}_i \MC_i$ for some invertible $t_i \times t_i$ matrix $\MC_i$. (For $i \in V_0$, $\MC_i$ is the identity matrix.) Set $\RMz'_i := \RMy\MM_i'$, so that $\RMz'_i = \RMy \widetilde{\MM}_i \MC_i= \RMz_i \MC_i$. If we write $\RMz'_i$ as $\bM \Rz'_{i,1}&\ldots&\Rz'_{i,t_i}\eM$, then $\RMz'_i := \RMy\MM_i'$ implies that
\begin{itemize}
\item[-] for $i \in V_0$, $\RMz_i'$ is a function of $\bM \Ry_1 &\ldots&\Ry_{\ell -1}\eM$ alone; and
\item[-] for $i \notin V_0$, only $\Rz'_{i,t_i}$ is of the form $\sum_{j=1}^\ell \alpha_j \Ry_j$, with $\alpha_{\ell} \ne 0$; all other components $\Rz'_{i,j}$, $j = 1,\ldots,t_i-1$, are functions of $\bM\Ry_1&\ldots&\Ry_{\ell -1}\eM$ alone.
\end{itemize}

For each $i \notin V_0$, consider the communication $\RMf_i$ from user $i$: write $\RMf_i$ as $\RMz_i \MA_i$ for a suitable matrix $A_i$. We then have
\begin{align*}
    \RMf_i = \RMz_i'\MC_i^{-1}\MA_i = \RMz'_i \MA_i',
\end{align*} 
with $\MA_i' := \MC_i^{-1}\MA_i$. Since the overall communication $\RMf_V$ consists of linear combinations of $\Ry_1,\ldots,\Ry_{\ell -1}$ alone, it must be the case that the last row of $\MA_i'$ is zero. Hence, $\RMf_i = \RMz_i^{\#} \MA_i^{\#}$, where $\RMz_i^{\#} := \bM\Rz'_{i,1}&\ldots&\Rz'_{i,t_i-1}\eM$, and $\MA_i^{\#}$ is $\MA_i'$ with the last row deleted. 
For $i \in V_0$, we simply set $\RMz_i^{\#} = \RMz_i'$ and $\MA_i^{\#}= \MA_i'$, so that we again have $\RMf_i = \RMz_i^{\#} \MA_i^{\#}$. 

For each $i \in V$, $\RMz_i^{\#}$ is a function only of $\bM \Ry_1&\ldots&\Ry_{\ell -1}\eM$, so that for $\RMz_V^{\#} := \bM \RMz_1^{\#}&\ldots&\RMz_m^{\#}\eM$, we have 
\begin{align*}
    H(\RMz_V^{\#}) \le H(\Ry_1,\ldots,\Ry_{\ell-1}) = (\ell-1) < H(\RMz_V).
\end{align*}
Note that $\RMz_V^{\#}$ is a finite linear source, since $\RMz_i^{\#} = \RMy\MM_i^{\#}$, where $\MM_i^{\#}$ is either equal to $\MM_i'$ (for $i \in V_0$) or is equal to$\MM_i'$ with its last row deleted (for $i \notin V_0$). Moreover, $\RMz_V^{\#}$ can be simulated by $\RMz_V$. Explicitly, for each $i \in V_0$, $\RMz_i^{\#}$ is equal to $\RMz_i\MC_i$ ($=\RMz_i'$), and for each $i \notin V_0$, $\RMz_i^{\#}$ is equal to $\RMz_i\MC_i$ ($=\RMz_i'$) with its last component deleted. Our arguments above show that the communication $\RMf_V = \RMz_V \MA$ is a non-interactive linear communication for $\RMz_V^{\#}$ as well. Finally, we argue below that the secret key $\RK$ can be generated from $(\RMz_i^{\#},\RMf_V)$ for each $i \in V$. Hence, $\rS(\RMz_V^{\#}) \le \rS(\RMz_V)$, which along with Lemma~\ref{lem:sim} proves the theorem.

For user $i$ to generate $\RK$, it must be a function of $(\RMz_i,\RMf_V)$. Equivalently, $\RK$ is a function of $(\RMz'_i,\RMf_V)$, since $\RMz'_i$ is an invertible function of $\RMz_i$. We claim that $\RK$ is in fact a function of $(\RMz_i^{\#},\RMf_V)$. This is trivially true for $i \in V_0$, so assume $i \notin V_0$. We say that a function $f(y_1,\ldots,y_n)$ is \emph{functionally dependent} on the variable $y_n$ if there exist $y_1',\ldots,y_n'$, and $y_n'' \ne y_n'$ such that $f(y_1',\ldots,y_{n-1}',y_n') \ne f(y_1',\ldots,y_{n-1}',y_n'')$. Note that $f(y_1,\ldots,y_n)$ is in fact a function of $y_1,\ldots,y_{n-1}$ alone iff it is not functionally dependent on $y_n$. We will argue that for $i \notin V_0$, $\RK$ is not functionally dependent on $\Rz'_{i,t_i}$.

Suppose, to the contrary, that $\RK$ is functionally dependent on $\Rz'_{i,t_i}$ for some $i \notin V_0$. Then there exist two distinct realizations, say, $\Mu$ and $\Mv$, of $(\RMz_i',\RMf_V)$ that differ only in the component $\Rz'_{i,t_i}$, which result in two distinct values of $\RK$: $\RK(\Mu) \ne \RK(\Mv)$. The realizations $\Mu$ and $\Mv$ share a common value of $(\RMz_i^{\#},\RMf_V)$, which is a function of $\bM \Ry_1&\ldots&\Ry_{\ell -1}\eM$ alone. Hence, there exists a realization of $\bM \Ry_1&\ldots&\Ry_{\ell -1}\eM$ that determines the realization of $(\RMz_i^{\#},\RMf_V)$ common to $\Ru$ and $\Rv$. Now, $Z'_{i,t_i} = \sum_{j=1}^\ell \alpha_j \Ry_j$ with $\alpha_\ell \ne 0$, so a given realization of $\bM \Ry_1&\ldots&\Ry_{\ell -1}\eM$ only suffices to determine the term $\sum_{j=1}^{\ell-1} \alpha_j \Ry_j$. Thus, we need two distinct realizations of $\Ry_{\ell}$, say, $\Ry'$ and $\Ry''$, to determine the values of the component $Z'_{i,t_i}$ in which $\Ru$ and $\Rv$ differ. In other words, $\Ru$ and $\Rv$ are determined, respectively, by $\bM \Ry_1&\ldots&\Ry_{\ell-1}&\Ry_\ell = \Ry'\eM$ and $\bM \Ry_1&\ldots&\Ry_{\ell-1}&\Ry_\ell = \Ry''\eM$ for some $\Ry' \ne \Ry''$. This makes $\RK$ functionally dependent on $\Ry_\ell$, which is impossible since, as seen earlier, $\RK$ is a function only of $\bM \Ry_1&\ldots&\Ry_{\ell -1}\eM$.This proves that $\RK$ cannot be functionally dependent on $\Rz'_{i,t_i}$, and hence,  $\RK$ is a  function of $(\RMz_i^{\#},\RMf_V)$, as desired. 

This proves that there exists a reduced source, $\RMz^{\#}_V$, of $\RMz_V$ obtained by linear processing with $\rS(\RMz^{\#}_V) = \rS(\RMz_V)$. 

\fi

\end{document}